\documentclass[conference]{IEEEtran}

\usepackage{multirow}
\usepackage{amsfonts}
\usepackage{amsmath}
\usepackage{cleveref}
\usepackage{color}
\usepackage{colortbl}
\usepackage{bm}
\usepackage{amssymb}
\usepackage{graphicx}
\usepackage{algorithmicx}
\usepackage{algpseudocode}
\usepackage{algorithm}
\usepackage{array}
\usepackage{fancyhdr}
\usepackage{epstopdf}
\usepackage{paralist}
\usepackage{url}
\usepackage{subfigure}
\usepackage[final]{changes}
\usepackage{xcolor}

\pdfoutput=1

\definecolor{Gray}{gray}{0.85}
\definecolor{LightGray}{gray}{0.95}
\newcolumntype{g}{>{\columncolor{Gray}}l}

\IEEEoverridecommandlockouts

\makeindex

\begin{document}
\IEEEoverridecommandlockouts
\IEEEpubid{\makebox[\columnwidth]{{978-1-4673-8463-6/16/\$31.00~\copyright~2016 IEEE. The IEEE copyright notice applies.} \hfill}
\hspace{\columnsep}\makebox[\columnwidth]{ }} 

\title{An IoT Architecture for Wide Area Measurement Systems: a Virtualized PMU Based Approach}

\author{\IEEEauthorblockN{Alessio Meloni, Paolo Attilio Pegoraro, Luigi Atzori and Sara Sulis}
\IEEEauthorblockA{DIEE, University of Cagliari, Italy\\
 \textit{\{alessio.meloni,paolo.pegoraro,l.atzori,sara.sulis\}@diee.unica.it}\\
}


}

\maketitle
\normalem

\begin{abstract}
Internet of Things (IoT) technologies are pervading different application domains by relying on sensing and actuating devices that share, process and present meaningful real-world information. One of the most important of these domains is certainly the Smart Grid (SG), where the use of advanced measurement and control equipment and the diffusion of communication technologies are making \added{an adaptive, reliable, and efficient management of the energy possible, with several new applications.} In this scenario, this paper focuses on one of the major IoT features, which is the virtualization, and proposes an IoT solution for wide area measurement systems where virtualized phasor measurement resources are introduced. Such a solution is intended to make a programmable and smart environment fostering interoperability, reusability and flexibility of SG services. 
\added{The performance of the system is evaluated for both the traffic generated and the latency to understand which scenarios can benefit 
from its deployment.}\\
\end{abstract}

\added{}


\section{Introduction}
In the last decade, the Internet of Things (IoT) \cite{atzori2014} paradigm has received increasing attention. Interest comes especially from those domains in which the creation of a cyber-physical world where everything can be probed, controlled, composed and updated allows the optimization of existing processes and the creation of new ones. 
One major building block behind the success of the IoT paradigm is the virtualization \cite{nitti2015}, in which counterparts in the cyber world are created for any real entity and its functionalities. As a matter of fact, several recent projects, which aim at creating a reference architecture for the IoT \cite{iota} \cite{icore} \cite{onem2m}, propose the abstraction from physical object's properties and processes. This is considered an important advancement towards interoperable and flexible IoT systems. Virtualization has the ability to: extend the functionalities, the features and the capabilities offered by entities and devices; make heterogeneous objects interoperable through the use of semantic descriptions; enable them to acquire, analyse and interpret information about their context in order to take relevant decisions; provide a programmable environment which fosters reusability, dynamic runtime configuration and flexibility.

These aspects are among those envisioned in the specific context of Smart Grids (SGs). The Smart Grid's Strategic Research Agenda \cite{sra2035} 
identified the need of a successful interfacing of heterogeneous grid equipment so as to ensure interoperability. Moreover, in the future it is likely that a greater flexibility for information flows and decision making will need to be introduced to satisfy the wide range of SG applications that go from high priority data for real-time control to lower priority data required for market purposes, asset management or planning activities. Last but not least, proposed solutions must be scalable (i.e. suitable for the needs ranging from those of small utilities to those of large utilities and industrial users).

In this paper, a SG architecture that includes the abovementioned virtualization functionalities to represent power system related components of various nature (smart meters, phasor measurement devices, data concentrators, circuit breakers and many more) is proposed and detailed for the case of a Wide Area Measurement System (WAMS) \added{\cite{terzija2011}}.  
WAMS is usually based on the data acquisition technology of Phasor Measurement Units (PMUs, \added{\cite{IEEE_Std_C37.118.1-2011}}) and 
the subsequent intermediate elaboration of Phasor Data Concentrators (PDCs), and allows for transmission system monitoring. However, in 
the SG context, the same WAMS concept can be extended and conceived in a wider sense as the integration of a set of heterogeneous 
technologies for remote and coordinated effective monitoring, providing a detailed and dynamic picture of the operating conditions of the 
grid. In this regard, PMUs are specifically considered here. \added{PMUs measure, in a synchronized way and with high rate, electrical 
quantities in different points of the network, thus enabling a meaningful view of the whole power system. However, the proposed analysis 
is not restricted to PMUs exclusive use in light of the wider meaning which can be attributed to the WAMS concept.}  

\added{In this regard, research activity is in act worldwide. In \cite{xin2011} and \cite{liakopoulos2012}, the concept design for a SG architecture leveraging virtualization was qualitatively discussed. \cite{maheshwari2013} discusses system state estimation in a private cloud, focusing on communication performance when sending data to different locations across the globe rather than giving a specific grid topology. \cite{chai2015} introduces the use of a topic-based publish-subscribe communication infrastructure.}

\added{In this work, a cloud-based monitoring system which leverages on a public cloud, complemented by distributed entities 
and specifying the use of REST APIs for communication among them, is presented. The proposed system allows both unicast 
communication (crucial for protection) and topic-based communication (crucial for scalability). Results, obtained also considering the 
standard 14-bus IEEE test system, are illustrated in details.} 

The paper is organized as follows: in Section \ref{sec:architecture} the general architecture proposed for the SG domain is presented and detailed for a WAMS 
using PMUs. Section \ref{sec:implementation} introduces the WAMS scenario adopted for simulations from both the electrical and the communication perspectives. In Section \ref{sec:results} the results obtained for the implemented network under various configurations are presented and discussed. Section \ref{sec:conclusions} summarizes the obtained results and outlines future work on the topic.

\section{Proposed Architecture}
\label{sec:architecture}

Depending on the specific scenario, \added{applications exist} with different interests in the PMU data: some might need just a subset of the provided information such as data at a low rate or averaged over a given time interval; others might need only some of the parameters of the \added{electrical} waveforms such as, for example, the frequency and the Rate Of Change Of Frequency (ROCOF). Moreover, the possibility to remotely control certain sensing or actuation properties related to devices is highly desirable in order to adapt the system based on contextual information under specific circumstances. Last but not least, the integration of real-time (i.e. critical) and non real-time applications is highly desirable for the sake of cost savings. In the following, an architecture capable of satisfying these functional requirements, \added {which details the general reference framework of iCore \cite{icore},} is presented. Fig. \ref{fig:arch} shows a logical diagram of the architecture.
 
\subsection{Abstraction}
\begin{figure}[t!]
	\centering
	\includegraphics[width=0.85\columnwidth]{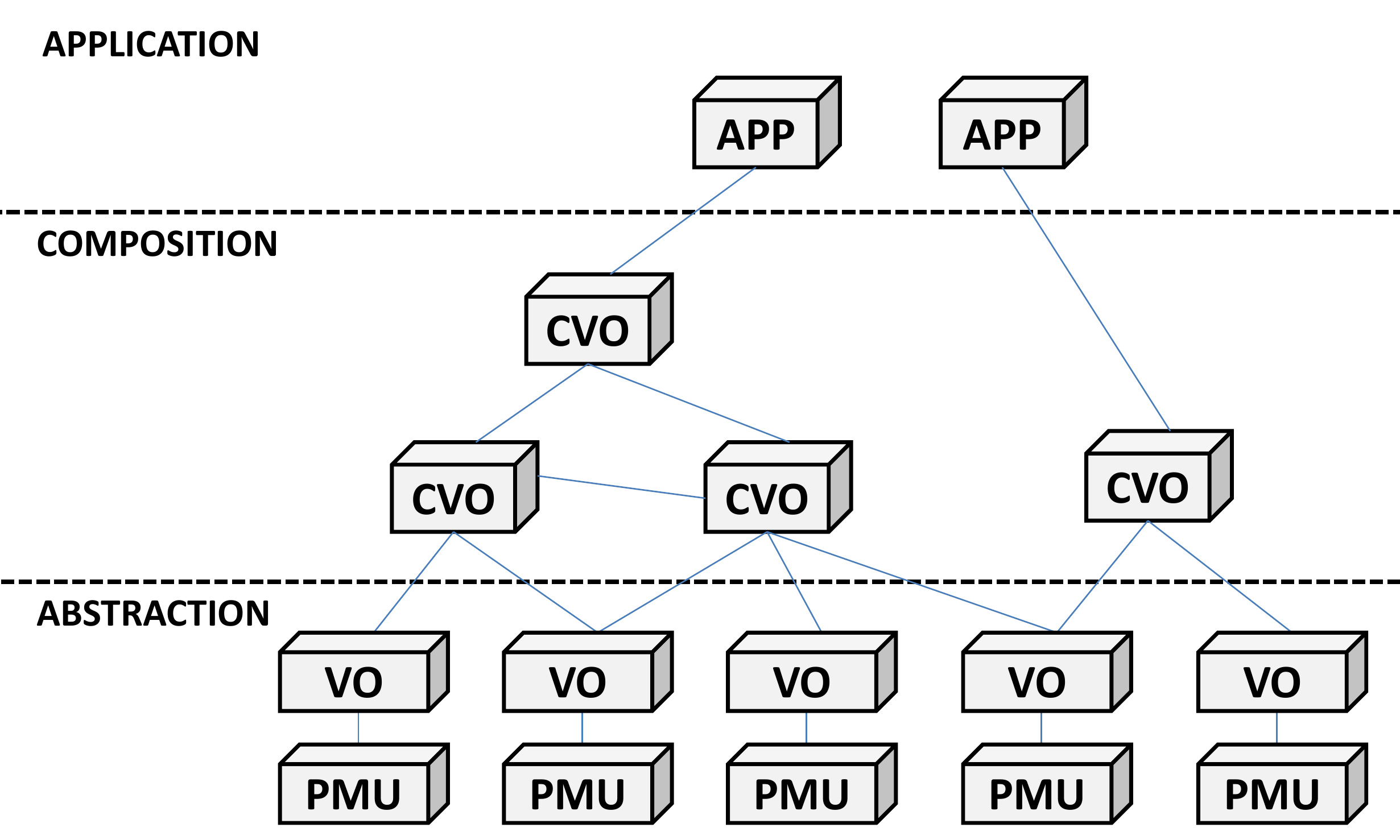}
	\caption{Proposed Architecture}
	\label{fig:arch}
	\vspace{-0.3cm}
\end{figure}

The first level is represented by the Abstraction layer. Virtualizing one or more physical objects such as PMUs and making them available as distributed resources with a programmable interface creates a more scalable and future-proof system. Currently, a commercial PMU is not required to be highly configurable
but just capable of understanding standard commands, such as the basic IEEE C37.118.2 ([10]) commands to turn on/off data frame
transmission of specific streams. For configuration, PMUs usually use different accesses and speak proprietary protocols and the system operator should be able to understand these as the PMUs don't have the power of using different languages. The introduction of a virtualization layer allows for overcoming these limitations creating a unified programmable framework for PMU output configuration where data is gathered by Virtual Objects (VOs) at the maximum data extent and possible rate. Moreover, data gathering takes place in a private dedicated subnetwork in which the communication traffic generated is predictable and highly efficient communication means, such as fiber optic cabling, can be used, in light of the short distances to cover and the affordable price. In particular, VOs can be thought as instances of a given template (similarly to the instances of classes in object-oriented programming languages), which can represent a single physical object as the PMU and its functionalities, as shown in Fig.\ref{fig:arch}.  

A VO abstracts the underlying technological heterogeneity and complexity of the physical objects, and makes the request of data (i.e. phasor measurements) more flexible by giving the possibility to interested entitites to receive only part of the measurements or data generated by PMU and elaborated by VOs with simple parsing or manipulation functions (e.g. average over a number of measurements). This service is made available both encapsulating the physical object communication standard (as for example IEEE C37.118.2 for PMUs) or using a common key-value semantic. Standard REST APIs interaction methods are used here for data exchange. It is thus possible to interact with a resource only knowing the Universal Resource Locator (URL) and using common HTTP methods (such as GET and POST) where the requester can indicate which data is of interest or set a trigger to automatically receive updates about given measurements. Moreover, using HTTP gives the possibility of an easy integration with security protocols such as SSL/TLS, which is an important aspect for most of applications.

\subsection{Composition}
Above the abstraction level of the proposed architecture are Composite Virtual Objects (CVOs). CVOs are cognitive mash-ups of semantically interoperable VOs created to aggregate VOs and accomplish higher level tasks. The CVO aim is to serve as mediator between abstracted functionalities and multiple high-level services which can reuse the same CVO. As an example, a CVO can be dedicated to the voltage or frequency data gathering of a given zone of the \added{electric} network for monitoring applications. Nevertheless, differently from PDCs which basically serve as data aggregators, multiple CVOs may also exist, with different functions in the same zone of the grid (e.g. to parse only certain data received from VOs or monitoring that certain thresholds are not exceeded). Another characteristic of CVOs is that they can be composed of other CVOs in a multi-tier structure and/or be linked to other CVOs of the same tier in order to accomplish a cooperative task (e.g. multi-area state estimation with a high-level controller implemented at the application level).

Differently from VOs, which are distributed on the edge of the communication network and close to the physical objects they are representing, CVOs can be placed locally or remotely with respect to the physical devices, depending on the needs and implemented functionalities. The advantages of having a CVO on the edge of the communication network are usually connected to: the smaller traffic load generated in the communication network; the diminished latency which provides a more responsive system for delay critical applications; the increased security ensured by the fact that CVO functions take place behind a firewall. The advantage of a remote CVO are: integration in a cloud infrastructure, which allows greatest computational and storage capabilities which elastically adapt to the dynamicity of the system without the need for operators intervention (e.g. because more computational power is needed for parsing measurement data); global visibility (differently from local CVOs), which facilitates the composition of services and the realization of applications.

An important property of CVOs is adaptability to grid events. In fact, CVOs must be able to adapt their mash-up of underlying VOs in a cognitive manner thus trying to guarantee a certain service to the application level. In other words, the use of entities which perform high level tasks should relieve applications from caring about the details connected to a certain function and allow CVOs to dynamically adapt to network changes and grid events independently, while trying to maintain the service level required by applications.  

Due to such characteristics, CVOs are the key module for many of the functionalities of the proposed architecture, as briefly illustrated in the following.

As a first example, the CVOs are perfect candidate for the PDC role, which is correlation and aggregation of PMU data. Usually, PDCs are organized in hierarchical structures in WAMS architectures. CVOs can obviously act as regional or primary PDCs, with additional advantages in terms of reconfigurability or computational capacity, which can be supported by cloud systems. The virtualization is pervasive, because it obviously allows the abstraction from rigidly defined concentrators architecture and assures scalability.

Commercial PMUs are usually equipped with a limited number of channels (frequently three for voltage and three for current measurements, corresponding to a three phase voltage and current pair) that allows only a partial monitoring of a given network node. Besides, in transmission networks, voltages and currents are often brought into different relay houses \cite{PMUGuide2007}.
For these reasons, vendors usually suggest to install more PMUs at a node, to cover all the converging branches, to limit both the wiring necessary to connect current transformers (CTs) and potential transformers (PTs) to the PMU and to optimize the communication infrastructure. It is important to recall that the device cost is just a small portion ($5\,\%$ or $10\,\%$, \cite{doePMUcosts2014}) of the overall cost of PMU installation and, thus, a proper definition of the communication architecture is directly related to the choice of the measurement system.
In the proposed architecture, a CVO can thus act, in cooperation with the VOs corresponding to the set of physical devices at a given network node, as a fully equipped, programmable, flexible, hardware-independent, multi-channel virtual PMU. Such virtual device can be integrated into the WAMS at different levels, thus limiting also the overhead of individual communication streams from the physical devices. As aforementioned, the communication protocol can be the IEEE C37.118.2 (as in a real PMU) or another one, depending on the interaction between CVO and applications.

Another interesting application of the CVO concept, is the possibility to design, in a decentralized way, a set of triggers or procedures based on PMU data (at different aggregation levels) able to rise alarms of events. For instance, simple real-time data manipulations can be performed locally for different applications, such as angle-based generation trip event detection, frequency-based line trip detection (see, for instance, the algorithm in \cite{2014PES_ZhoLiuDon_Freqbasedlinetrip}) or dynamic line rating.

The interface between VOs and CVOs, as outlined in the previous subsection, is made of two different communication patterns depending on the specific application requirements. Data from the VO can be pulled by the CVO using the HTTP GET method; alternatively, for those CVOs that need the data when a specific event happens or at a pre-defined frequency, data can be forwarded setting a trigger on the VO (e.g. ``send data at the beginning of each minute") so that it will send data in push mode to the CVO using the HTTP POST method.

\subsection{Application}
At the highest level of the proposed layered architecture are the applications, which are responsible for gathering high-level and elaborated information from one or more CVOs and running the needed application services. For example, among the others, frequency monitoring, visualization services for different quantities, such as the power flowing in a given number of branches of the network or the voltages at critical nodes, can be considered.

In the wider SG perspective, one of the most important applications is the state estimation (that is the monitoring basis for many other applications). The proposed virtualization architecture can be seen as a definition of a flexible monitoring system, which can be exploited to integrate measurement data gathered by other instruments (such as conventional devices equipped with an adapter, smart metering devices) and other information sources, such as the state of switches or reclosures, tap changer positions, and whatever can be helpful for the  state estimation \cite{2014Energycon_PauPegSul_PMUinSE}. Such application can also operate with different policies (in hierarchical or decentralized way, for instance), involving both the application and composition logical levels.

\subsection{Topic-based interaction}
\begin{figure}[htbp!]
	\vspace{-0.3cm}
	\centering
	\includegraphics[width=0.5 \columnwidth]{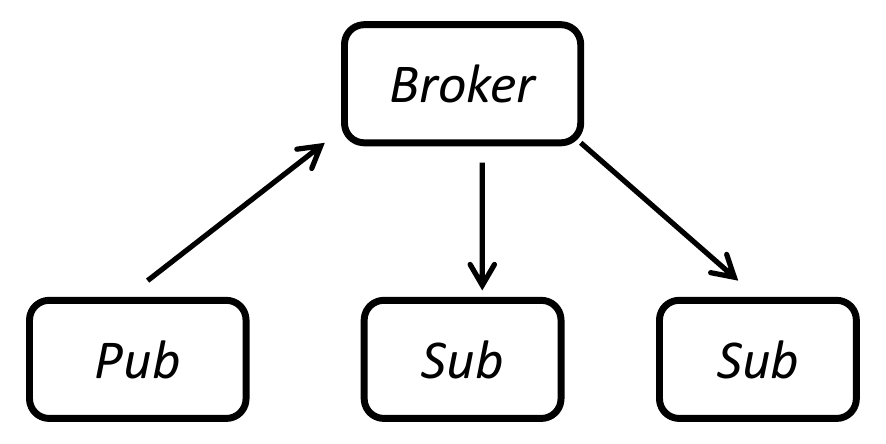}
	\caption{Publish-subscribe pattern}
	\label{fig:pubsub}
\end{figure}

So far, we have analyzed VOs as the interface of physical objects to the cyber world. Nevertheless, also the opposite is true. VOs function as interfaces for actuation and control commands sent from the highest levels of the architecture to the underlying building blocks. This can either be accomplished using REST APIs for unicast communication or leveraging on the advantages of the publish-subscribe paradigm for multicast communications. In the latter case, communication between entities takes place in a topic-based manner in which interested parties can subscribe to a given topic or publish information on that topic leveraging on a external broker which manages the underlying communications (Fig. \ref{fig:pubsub}). This allows for setting up a scalable system where multiple entities are advertised in an automated way. As an example, if multiple VOs report to a common CVO the frequency and the ROCOF measured in certain nodes of the grid, they could be subscribed to a specific topic. In case the CVO senses that one of the considered PMUs has measured a ROCOF exceeding a certain threshold, it could simply ask VOs to send more frequent updates about the measured frequency by just advertising on the selected topic.

\begin{figure}[htbp!]
	\vspace{-0.3cm}
	\centering
	\includegraphics[width=0.9 \columnwidth]{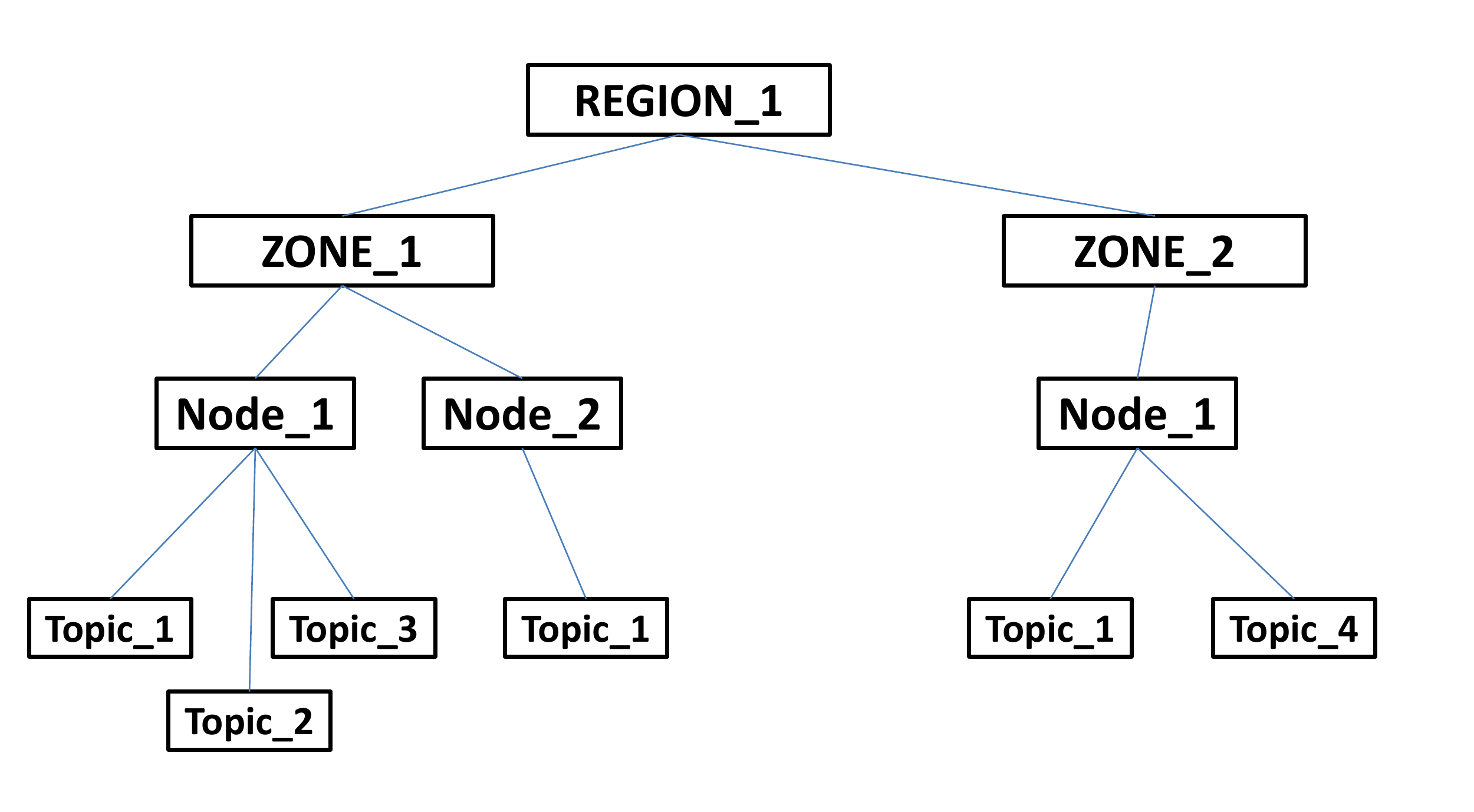}
 	\vspace{-0.3cm}
	\caption{Example of MQTT topic tree}
	\label{fig:mqtt}
\end{figure}

In our architecture, the use of the MQTT protocol is considered. In MQTT, topics are organized in a tree structure (see Fig. \ref{fig:mqtt}). To select a given topic, the tree can be explored using a subfolder-like nomenclature. For example, the topic ``REGION\_1/ZONE\_1/Node\_2/Topic\_1" can be chosen. Moreover, two wildcards can be used:
\begin{itemize}
\item the character \# allows to select all the subtopic branches under a certain topic (e.g. ``REGION\_1/ZONE\_1/\#" identifies all the zones of ZONE\_1 and the underlying topics);
\item the character + can be used to identify a topic with the same name under different branches (e.g.``REGION\_1/+/+/Topic\_1" identifies all the topics ``Topic\_1" under different branches which could be for example the frequency).   
\end{itemize}

\section{Implementation of the Proposed Architecture in the Considered WAMS Scenario}
\label{sec:implementation}

In this section, the proposed architecture is explained in details in a test case. The aim is to have a practical application field where the concepts described in the previous sections can be translated in a real context. The implementation is illustrated and the assumptions made for the tests are briefly recalled.

\subsection{Proposed Scenario}
\begin{figure}[!t]
\centering
\includegraphics[width=0.85\columnwidth, trim=0 220 0 200 ]{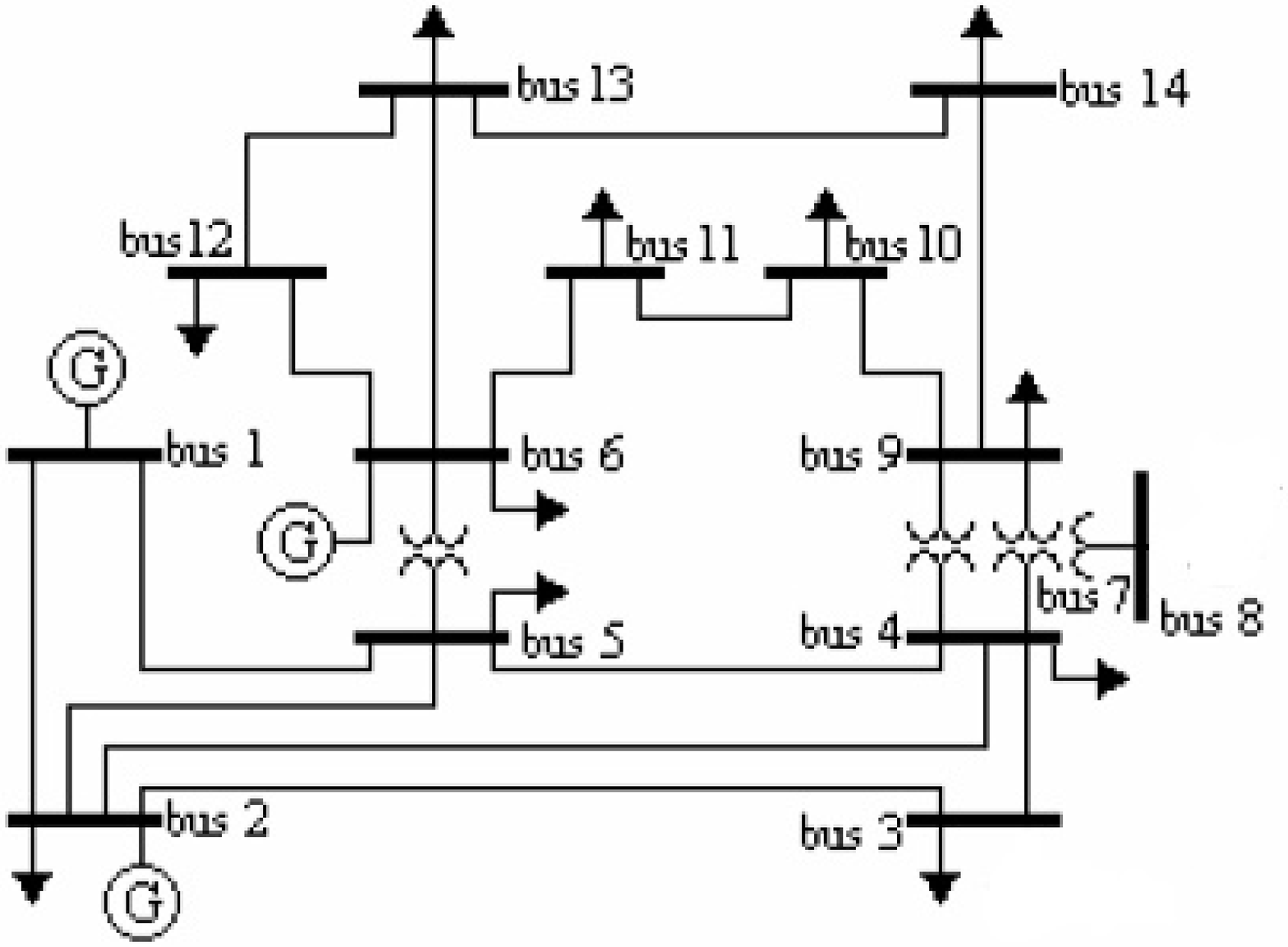}
\caption{IEEE 14-bus test system.}
\label{fig:14-bus}
\vspace{-0.3cm}
\end{figure}

As an example of the application of the proposed architecture, an implementation of the monitoring system for the IEEE 14-bus benchmark network has been considered for the tests (Fig. \ref{fig:14-bus}). 
Details about the network (line parameters, nominal loads and generators) can be found in \cite{TestCaseArchive}. However, for the aim of the communication tests in the following, only the topology of the network is needed and the network is used to define a realistic measurement system. 
The emulation of the PMU-based monitoring system is performed by means of real PMU prototypes developed using National Instruments CompactRIO modular technology. Each prototype is synchronized by means of a GPS receiver and can work both as a fully equipped PMU on acquired signals and as a PMU emulator, that is a synchronized device that can measure pre-stored signals or simulate an expected measurement output.

A measurement configuration optimal in terms of minimum number of monitored nodes, while keeping observability (see \cite{XuAburRepPMU} for details), is adopted for the tests. The nodes 2, 6, 7 and 9 are chosen to be monitored. In particular, the voltage of the node and all the currents of the incident branches are measured.
Each node can be monitored by more PMUs, as aforementioned, and, in this discussion, a PMU is assumed to measure two quantities (one voltage and one current or two currents, in three-phase configuration). With such assumption, nodes 2, 6 and 9 require three PMUs while two PMUs are placed at node 7.

\subsection{Communication overlay}
The implementation of the proposed communication overlay has been performed as follows. Locally, virtualized entities such as VOs and CVOs are implemented in dedicated devices capable of maintaining one or more TCP connections with PMUs and connected to the rest of the subnetwork and the Internet in order to send gathered data using REST APIs. Ethernet cabling is supposed in a local subnetwork, which is currently one of the solutions used in substations. For the tests, the local devices have been placed in the laboratories of the University of Cagliari in Italy.  

Remote CVOs and applications have been implemented using Google App Engines, which is a cloud-based Platform as a Service (PaaS) providing dedicated resources for customized implementations in which the necessary computational and storage needs are managed elastically (grown or shrunk) according to the needs. In particular, the considered CVOs and applications have been placed in the closest farm for European users which is located in St. Ghislain, Belgium.

\begin{figure}[htbp!]
 \centering
 \subfigure[Local CVO]
   {\includegraphics[width=0.49 \columnwidth]{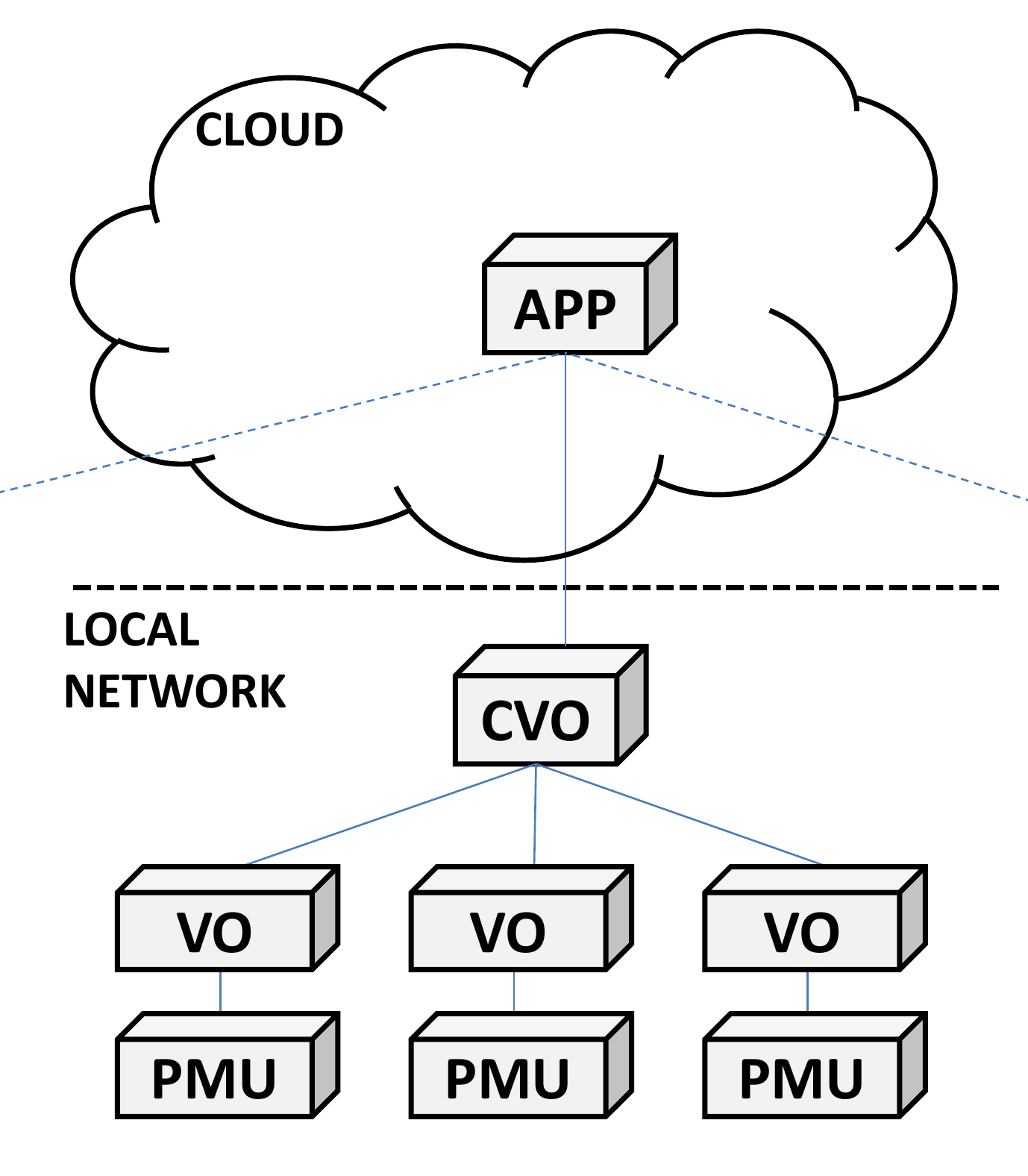}\label{fig:localCVO}}
 \subfigure[Remote CVO in the cloud]
   {\includegraphics[width=0.49 \columnwidth]{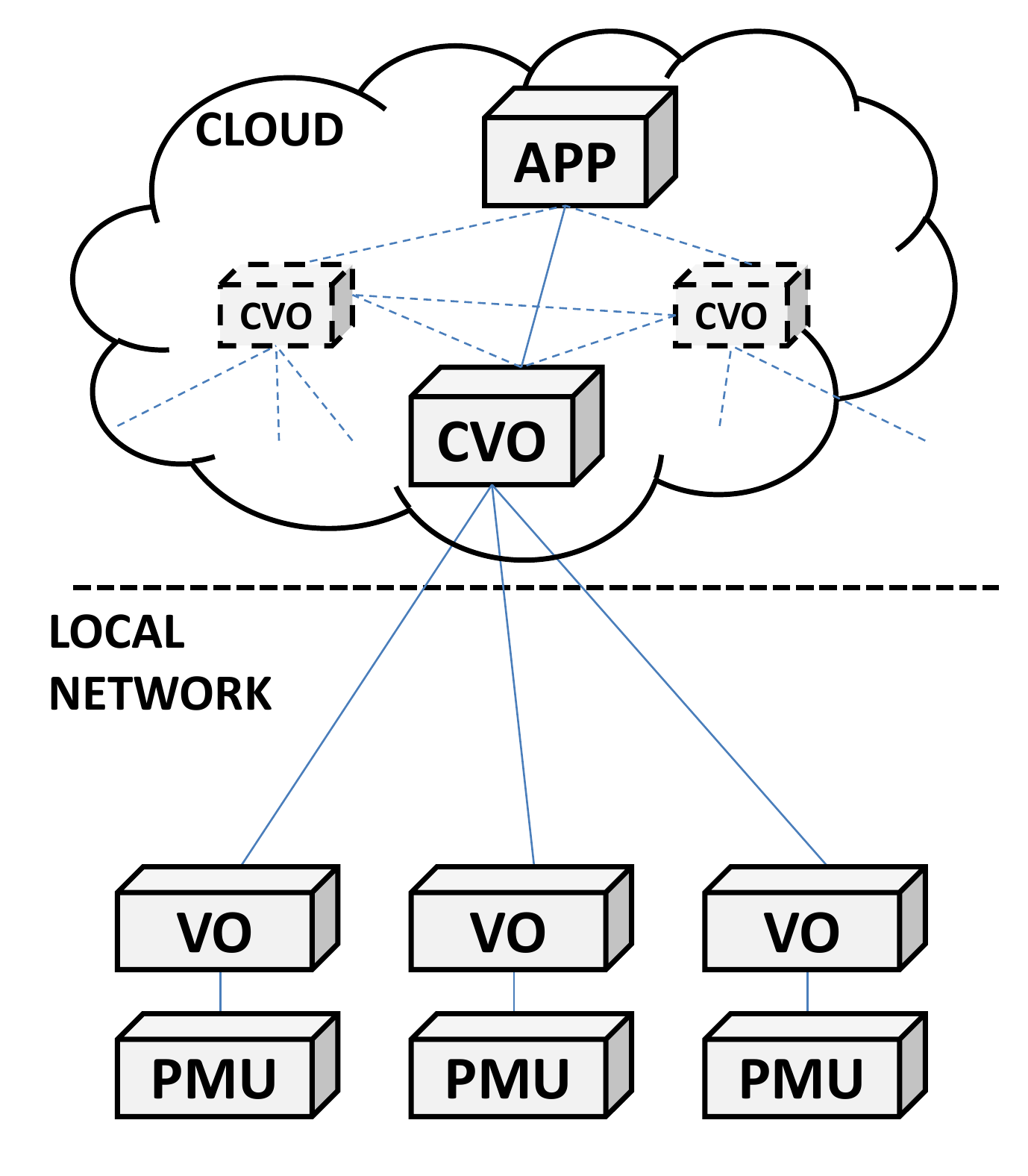}\label{fig:remoteCVO}}
 \caption{CVO placement options.}
  \end{figure}

The subsequent simulations consider two different cases: the case in which CVOs concentrate data gathered from VOs locally and then send aggregated data to the cloud (Fig. \ref{fig:localCVO}); the case in which the CVO is implemented in the cloud and receives one flow per physical PMU (Fig. \ref{fig:remoteCVO}). The goal is to discuss advantages and disadvantages of the former versus the latter approach, taking into account that real implementations will certainly have a blend of the two solutions in light of their characteristics.

\section{Results}
\label{sec:results}

The two schemes shown in Figs. \ref{fig:localCVO} and \ref{fig:remoteCVO} correspond to the case of three physical PMUs at a single node (node 2 in Fig. \ref{fig:14-bus}, for instance) and represent the decentralized and centralized approaches to node virtualization, respectively.
In the former case, the virtualization allows for having a logical device represented by the CVO, which serves as an aggregator for the higher layers of the architecture before the local network is left. The distributed approach allows for considerably saving the bandwidth used, as can be seen from Table \ref{table:bandwidth}. In the latter case, the same function is done remotely, which surely requires more bandwidth but allows a finer data gathering from single VOs representing physical PMUs. The Table reports the required bandwidth for the implemented HTTP communication from node 2 to the cloud in case of local and remote CVO, when different PMU data frame configurations are considered, in order to highlight the bandwidth-related advantages of a CVO which processes data locally. In particular, two data set are presented: Config A (6 phasors, two 3-phase channels), which is a typical PMU configuration and Config B (12 phasors, two 3-phase channels with positive, negative and zero sequences), which is a redundant configuration sometimes chosen by transmission system operator (TSO). Config A and B are evaluated for both fixed 16-bit and floating-point formats to give lower and upper limits for each configuration. Halfway configurations are often used, such as floating point format for phasors and fixed 16-bit format for frequency and ROCOF. The bandwidth usage in this case is then in-between the results obtained for the two presented configurations.
The savings in the tests are always above $45\,\%$. Nevertheless, it should be noticed that these results also depend on the specific HTTP implementation and the considered headers. For this reason, a loose lower
bound has also been computed considering only the PMU payload plus 40B of the TCP-IP packet header. For Config A with floating-point format, for example, this means a bandwidth saving of $32.7 \%$.

\begin{table}[!htbp]
\renewcommand{\arraystretch}{2}
\centering
\caption{Node communication bandwidth for the decentralized/centralized approaches with different data configurations}
\begin{tabular}{l|c|c|c}
\hline
\multirow{2}{*}{\bfseries PMU Phasors Num} & \multicolumn{2}{c|}{\bfseries Bandwidth [bps]} & \multirow{2}{*}{\bfseries saving [\%]}\\
\cline{2-3}
& {\bfseries Local CVO} & {\bfseries Remote CVO} &\\
\hline
Config A - fixed & 126400 & 307200 & 58.9 \\
Config A - floating & 160000 & 340800 & 53.1 \\
Config B - fixed & 155200 & 336000 & 53.8 \\
Config B - floating & 217600 & 398400 & 45.4 \\
\hline
\end{tabular}
\label{table:bandwidth}
\end{table}

Latency has also been analyzed, which has been measured on $2500$ trials as the interval between the timestamp of the PMU packets and the moment in which data regarding those packets is received by the application level, which happens after the CVO receives all the expected packets with the same timestamp from VOs.   

\begin{figure}[b!]
	\centering
	\includegraphics[width=0.95 \columnwidth]{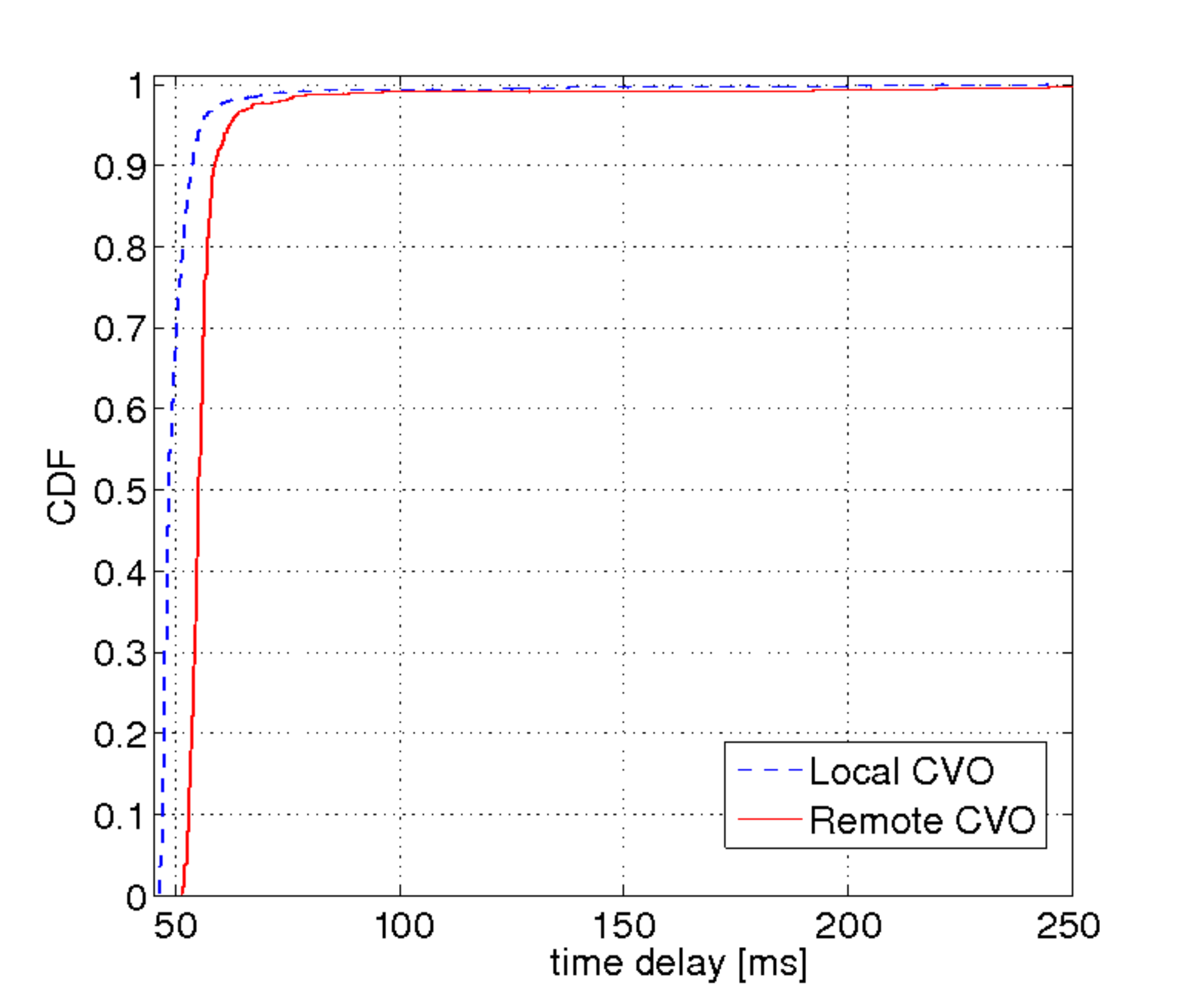}
	\caption{Latency in terms of cumulative distribution function}
	\label{fig:cdf}
	\vspace{-0.3cm}
\end{figure}

As shown in Fig. \ref{fig:cdf}, using a local CVO performs better in terms of latency. In fact, establishing multiple connections with a remote entity which is reached over the Internet puts more uncertainty on the delay between received packets, which is not the case for a local network in which the channel is generally faster and a proper communication's Quality of Service (QoS) can be ensured. From the figure it can be noted that 50 msec is the maximum latency for almost 70\% of packets with the local approach, which is a target never achieved in the other case. Table \ref{table:latency} shows the results for significant characteristics of the latency. Both in the case of a local and a remote CVO, the $500$ ms threshold is satisfied with a $100\,\%$ dependability, which translates to the possibility to use the described system even for remote services, such as, sending remote operator commands and events/alarms that are currently integrated in a poorly interoperable and not future-proof SCADA system. For a $100$ ms dependability, local CVOs behave slightly better while it can be seen that for really stringent latency constraints, only a local CVO guarantees a good dependability (approximately two-thirds). This is an interesting result for future optimization of the proposed system in which the needs posed by fast automatic interactions such as those regarding protection are satisfied in a IoT-based platform.


\begin{table}[t!]
\renewcommand{\arraystretch}{2}
\centering
\caption{Minimum, mean and maximum latency and dependability for the case of local and remote CVO.}
\begin{tabular}{|g|c|c|}
\hline
\rowcolor{LightGray}
{\bfseries Index} & {\bfseries Local CVO} & {\bfseries Remote CVO}\\
\hline
Minimum Latency & 46 ms & 51 ms\\
Mean Latency & 50 ms & 58 ms\\
Maximum Latency & 245 ms & 264 ms\\
\hline
50 ms Dependability & 66\% & 0\%\\
100 ms Dependability & 99.5\% & 99\%\\
500 ms Dependability & 100\% & 100\%\\
\hline
\end{tabular}
\label{table:latency}
\end{table}

Finally, tests considering the state estimation application running in the cloud have been considered
to simulate the whole monitoring platform. In particular, a linear state estimator based on a node voltage state in rectangular coordinates has been tested, relying on the network data in \cite{TestCaseArchive} and on the described PMU-based measurement system. An average computation time of $5.0$ ms with a standard deviation of $0.6$ ms has been found on 2500 trials. Thank to the cloud computing capabilities, the computation can thus have a lower impact on the overall monitoring system latency with respect to the communication, thus underlining the importance of a proper architectural design.

\section{Conclusions}
\label{sec:conclusions}
The paper presents an IoT architecture for the monitoring of power systems. In particular, the possibility to exploit the virtualization capability of IoT, along with the computation power and flexibility of the cloud, to design a PMU-based wide area measurement system is discussed. 

The results show how the definition of a proper architecture allows  limiting the bandwidth and latency of the communication, while building a flexible framework for the development of applications.
The potentialities of the proposed architecture are noteworthy and can prepare the way for a decentralized approach to network monitoring.


\section*{Acknowledgement}
This work has been supported by Regione Autonoma della Sardegna, L.R. 7/2007: ``Promozione della ricerca scientifica e dell'innovazione 
tecnologica in Sardegna, annualit\`{a} 2012, CRP-60511".
\bibliographystyle{IEEEtran}
\bibliography{IEEEabrv,references}

\end{document}